\documentclass{jfm}

\usepackage{graphicx}
\usepackage{natbib}

\usepackage{amsmath}
\usepackage{amssymb}
\usepackage{color}

\ifCUPmtlplainloaded \else
  \checkfont{eurm10}
  \iffontfound
    \IfFileExists{upmath.sty}
      {\typeout{^^JFound AMS Euler Roman fonts on the system,
                   using the 'upmath' package.^^J}%
       \usepackage{upmath}}
      {\typeout{^^JFound AMS Euler Roman fonts on the system, but you
                   dont seem to have the}%
       \typeout{'upmath' package installed. JFM.cls can take advantage
                 of these fonts,^^Jif you use 'upmath' package.^^J}%
       \providecommand\upi{\upi}%
      }
  \else
    \providecommand\upi{\upi}%
  \fi
\fi

\ifCUPmtlplainloaded \else
  \checkfont{msam10}
  \iffontfound
    \IfFileExists{amssymb.sty}
      {\typeout{^^JFound AMS Symbol fonts on the system, using the
                'amssymb' package.^^J}%
       \usepackage{amssymb}%
         
         \let\geq=\geqslant
      }{}
  \fi
\fi

\ifCUPmtlplainloaded \else
  \IfFileExists{amsbsy.sty}
    {\typeout{^^JFound the 'amsbsy' package on the system, using it.^^J}%
     \usepackage{amsbsy}}
    {\providecommand\boldsymbol[1]{\mbox{\boldmath $##1$}}}
\fi

\definecolor{red}{rgb}{0,0,0}
\definecolor{black}{rgb}{0,0,0}

\newcommand{\be}{\begin{equation}}
\newcommand{\ee}{\end{equation}}

\newcommand{\bs}{\begin{subequations}}
\newcommand{\es}{\end{subequations}}

\newcommand{\ba}{\begin{array}{c}}
\newcommand{\ea}{\end{array}}

\newcommand{\bd}[1]{{\boldsymbol #1}}

\newcommand{\br}{\bd{r}}
\newcommand{\bk}{\bd{k}}
\newcommand{\bv}{\bd{v}}

\newcommand{\bn}{\bd{n}}
\newcommand{\bq}{\bd{q}}
\renewcommand{\bv}{\hat{\bd{v}}}
\newcommand{\bx}{{\boldsymbol\tr}}
\newcommand{\bdcdot}{\bd{\cdot}}
\newcommand{\bOm}{\bd{\Omega}}

\newcommand{\tzeta}{\tilde{\zeta}}
\newcommand{\tz}{\tilde{z}}
\newcommand{\tx}{\tilde{x}}
\newcommand{\ty}{\tilde{y}}
\newcommand{\tr}{\tilde{r}}

\newcommand{\ET}{{E\&T}}

\newcommand{\rmi}{\mathrm{i}}
\newcommand{\rmd}{\mathrm{d}}

\newcommand{\rme}{\mathrm{e}}

\newcommand{\zff}{\tzeta_\text{f.f.}}

\newcommand{\half}{{\textstyle \frac1{2}}}

\DeclareMathOperator{\sign}{sign}
\DeclareMathOperator*{\Res}{Res}
\DeclareMathOperator*{\Min}{min}

\title[Oscillating point source in shear current]{Waves from an oscillating point source with a free surface in the presence of a shear current}
\author[S. \AA. Ellingsen and P. A. Tyvand]{Simen \AA. Ellingsen$^1$\thanks{Email address for correspondence: simen.a.ellingsen@ntnu.no} and Peder A. Tyvand$^2$}
\affiliation{$^1$Department of Energy and Process Engineering, Norwegian University of Science and Technology, N-7491 Trondheim, Norway \\
$^2$Department of Mathematical Sciences and Technology, Norwegian University of Life Sciences, N-1432 \AA s, Norway}

\begin{document}

\maketitle
\begin{abstract}
We investigate analytically the linearized water wave radiation problem for an oscillating submerged point source in an inviscid shear flow with a free surface.
A constant depth is taken into account and the shear flow increases linearly with depth. The surface velocity relative to the source is taken to be zero, so that Doppler effects are absent. We solve the linearized Euler equations to calculate the resulting wave field as well as its far-field asymptotics. For values of the Froude number $F^2=\omega^2 D/g$ ($\omega$: oscillation frequency, $D$ submergence depth) below a resonant value $F^2_\text{res}$ the wave field splits cleanly into separate contributions from regular dispersive propagating waves and non-dispersive ``critical waves'' resulting from a critical layer-like street of flow structures directly downstream of the source. In the sub-resonant regime the regular waves behave like sheared ring waves while the critical layer wave forms a street of a constant width of order $D\sqrt{S/\omega}$ ($S$ is the shear flow vorticity) and is convected downstream at the fluid velocity at the depth of the source. When the Froude number approaches its resonant value, the the downstream critical and regular waves resonate, producing a train of waves of linearly increasing amplitude contained within a downstream wedge. 
\end{abstract}

%%%%%%%%%%%%%%%%%%%%%%%%%%%%%%%%%%%%%%%%%%%%%%%%%
%%%%%%%%%%%%%%%% S E C T I O N %%%%%%%%%%%%%%%%%%
%%%%%%%%%%%%%%%%%%%%%%%%%%%%%%%%%%%%%%%%%%%%%%%%%
\section{Introduction}

The submerged oscillatory point source is pivotal for three-dimensional water waves with a localized cause. \cite{kochin39,kochin40} gave the first mathematical solutions where the source satisfies the linearized free-surface conditions and radiation conditions at infinity. \cite{wehausen60} summarized the classical solutions for submerged oscillatory sources. They are useful as Green functions for solving diffraction and radiation problems in marine hydrodynamics \citep{newman77,faltinsen90}.
The description of submerged bodies in inviscid flow using flow singularities rather than a full boundary value solution is highly economical, and has been hugely successful for irrotational flow.
No corresponding Green function theory exists when a shear current is present, however, and as a first step in this direction we consider the properties of the basic building brick for such a theory: the submerged oscillating point source.

In the preceding paper \citep[][abbreviated \ET\ hereafter]{ellingsen15a} we showed how the standard theory of the oscillating line source in 2D with free surface and uniform shear based on potential theory is inadequate. Potential theory has traditionally been applied to this system on the grounds that when vorticity is constant the wave motion will be irrotational due to conservation of circulation by Kelvin's circulation theorem. The 2D radiation problem was solved under this assumption by 
\citet{tyvand14,tyvand15} with both zero and nonzero surface velocity relative to the oscillating source, respectively. However, we recently discovered that the violation of Laplace's equation at the source's position itself, it is necessary to treat the problem with the full Euler equations (see \ET). The resulting wave pattern obtains a correction from the formation of a 
{\color{black}
critical layer-like train of downstream vortices 
}
at the source's depth which can completely dominate the wave picture in certain parameter regimes, and drastically alters the wave picture downstream of the source (as seen from a system where source and undisturbed surface are at rest). 
The existence of critical layer type solutions in 2D and uniform shear had previously been pointed out in the mathematical literature \citep{ehrnstrom08,wahlen09}, confirming an age old conjecture by Lord Kelvin \citep{kelvin1880}. The problem is a classic one and has attracted considerable recent attention; see \cite{constantin11} for a review.

In the present paper we will extend the work in 2D from \ET\ to three dimensions, still using the simplifying assumption that the source is at rest relative to the surface of the fluid. While a restriction which we will want to lift in the future, this assumption is a significant simplification because complicating Doppler effects are eschewed, allowing us to study several novel features unobscured by additional complexity.

In 3D and in the presence of shear current, potential theory is certainly not an option \citep[e.g.,][]{ellingsen14c}. Possibly for this reason the literature on 3D water waves with vorticity is small; a ring wave theory by \cite{johnson90}, 
{\color{red}
and some works in an oceanographic context, e.g., \citet{shrira93} and \citet{mellor03}, 
are notable examples. Progress was made recently, when it was realised by one of us 
}
that a simple solution exists to the linearised Euler equation in 3D in the presence of uniform shear, and the solution was used to solve the linear ship wave problem \citep{ellingsen14a}, as was the corresponding Euler--Cauchy problem of an initial disturbance \citep{ellingsen14b}. Both efforts have been generalised and expanded by \citep{li16b,li15a}. The present theory follows essentially the same procedure with the only significant difference being the presence of the source making the continuity equation inhomogeneous. %That such a solution should exist is perhaps not so surprising in retrospect, considering the general considerations of generalising 2D results to oblique angles by \citet{benney66}. 

In a famous review, \citet{peregrine76} names several situations where the depth dependence of a current will affect surface waves. Noting that shear is typically present either near the surface (for example due to wind shear as considered by, e.g., \citet{shemdin72,jones01}) or in a boundary layer near the bottom \citep[e.g.,][]{soulsby93}, and that a wave can ``see'' approximately half its wavelength into the deep, the presence of vorticity will affect waves whose wavelength is either sufficiently short to be affected strongly by the surface layer, or so long as to be affected by the entire water column. Situations where shear may be significant also include tidal currents \citep[e.g.][]{dyer71}. The Green functions summarised in the famous review of \citet{wehausen60} are all for irrotational flow. To our knowledge it still holds true today that the literature on fluid mechanical Green's functions is restricted to potential flow solutions. Hence, no valid Green function theory exists for floating bodies in shear currents, and the current effort takes perhaps the first step in this direction.

In comparing the 2D result of \ET\ with the 3D treatment herein, note that while the background flow in our 3D geometry is presumed to posess a constant vorticity, this is not the case for the flow as a whole. It was shown by \cite{constantin11b} and others that an assumption of constant nonzero vorticity implies that the flow is strictly 2D. This apparent paradox was recently addressed by \cite{ellingsen16}, demonstrating that a plane wave propagating at an oblique angle with a sub-surface shear flow of uniform vorticity will shift the vortex lines of the basic flow and be, in an Eulerian sense, rotational, and the total fluid motion has non-constant vorticity. Thus there is no conflict between Constantin's theorem and the present work. 

%%%%%%%%%%%%%%%%%%%%%%%%%%%%%%%%%%%%%%%%%%%%%%%%%
%%%%%%%%%%%%%%%% S E C T I O N %%%%%%%%%%%%%%%%%%
%%%%%%%%%%%%%%%%%%%%%%%%%%%%%%%%%%%%%%%%%%%%%%%%%
\section{Mathematical model}

\begin{figure}
  \begin{center}
  \includegraphics[width=.7\textwidth]{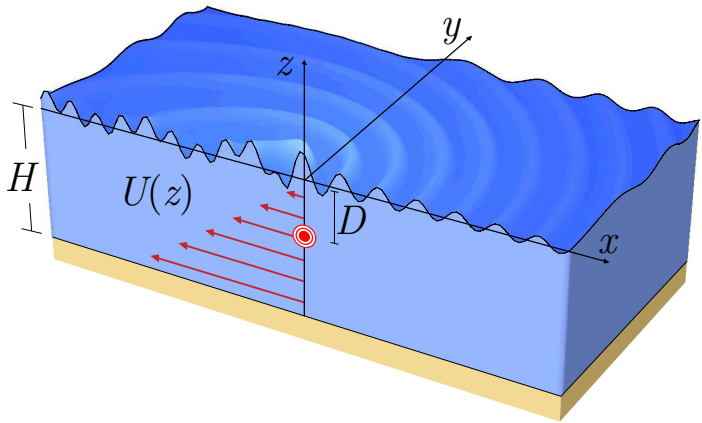}
  \end{center}
  \caption{The geometry considered: an oscillating point source sits at depth $D$ in a uniform shear current so that the source is at rest relative to the undisturbed water surface.}
  \label{fig:geom}
\end{figure}

We consider an inviscid and incompressible fluid in a steady shear flow, where the shear flow is aligned with a horizontal $x$ axis. 
The fluid has constant depth and a free surface subject to constant
atmospheric pressure. Cartesian coordinates $x, y, z$ are introduced,
where the $z$ axis is directed upwards in the gravity field and the
$x, y$ plane represents the undisturbed free surface. The gravitational
acceleration is $g$, and $\rho$ denotes the constant fluid
density. 
We neglect surface tension throughout. 
The velocity perturbation vector is denoted by $\bv=(\hat{u},\hat{v},\hat{w})$, and the full pressure is $P = -gz + \hat{p}$ where $\hat p$ is the perturbation.
The surface elevation is denoted by $\zeta(x,y,t)$, and the overall problem is sketched in 
Fig.~\ref{fig:geom}.
There is a wave motion driven by a fixed oscillating point source located in the point $(0,0,-D)$. The water wave problem will be linearized with respect to the surface elevation and the velocity 
and pressure perturbations.

We assume constant fluid depth $H$. There is a basic horizontal shear flow $U(z)$ in the $x$ direction
\begin{equation}\label{eq:shearVel}
  U(z) = S z,~~z < \zeta.
\end{equation}
The basic flow thus contains a uniform vorticity $S$ in the $y$ direction, whose influence on the suface motion is our principal concern.

Euler's equation of motion can be written
\begin{equation}\label{eq:euler}
  \bd{a} = - \frac{1}{\rho} \nabla P - g \bd{e}_z,
\end{equation}
where $\bd{a}$ is the acceleration vector and $\bd{e}_i, i\in\{x,y,z\}$ are the Cartesian unit vectors. 

The linearized kinematic free-surface condition is
%simple when no surface velocity is present,
\begin{equation}\label{eq:surfkincond}
\left. \hat{w} \right\vert_{z=0} = \zeta_t ,
\end{equation}
with subscripts denoting partial derivatives.

In the absence of both surface tension and viscosity the dynamic boundary condition is simply that pressure be constant (chosen equal to zero) on the free surface. This may be written by means of the tangential component of the Euler equation,
\begin{equation}\label{eq:surfcond}
  \left[\bd{a} - (\bd{a}\bdcdot\bd{n})\bd{n} + g \bd{e}_z - g (\bd{e}_z\bdcdot\bd{n})\bd{n}\right]|_{z=\zeta} = 0,
\end{equation}
because $\nabla P-\bn(\bn\bdcdot\nabla P) = 0$ there. 
Here the surface normal vector is denoted by 
$\bd{n}$. 
According to linear theory, the surface normal is given by 
$\bd{n}=\bd{k}-\nabla \zeta$. 
We linearize this dynamic free-surface condition and decompose it in the $x$ and $y$ direction, giving
\bs
\begin{align}\label{eq:surfcond1}
  (\hat{u}_t + S \hat{w})|_{z=0} =& - g \zeta_x,\\
  \label{eq:surfcond2}
  \hat{v}_t|_{z=0}   =& - g \zeta_y,
\end{align}
\es
By combining these two dynamic conditions we get
\begin{equation}\label{eq:surfcond3}
  (\hat{u}_{xt} + \hat{v}_{yt} + S \hat{w}_x)|_{z=0} =  -g (\zeta_{xx}+\zeta_{yy}),
\end{equation}
and 
use the incompressibility of the flow to
eliminate the horizontal velocities to get
\begin{equation}\label{eq:dynamic}
  (\hat{w}_{zt} - S \hat{w}_x)|_{z=0} =  g (\zeta_{xx}+\zeta_{yy}).
\end{equation}
The dynamic free-surface condition is thus expressed in terms of the vertical velocity and the elevation.

The last boundary condition is the bottom condition
\begin{equation}\label{eq:bottom}
\hat{w} = 0,~~z=-H.
\end{equation}
We will work with finite $H$, giving special attention to the limit $H \rightarrow \infty$.

We assume time-periodic flow with given angular frequency $\omega$. This
flow is driven by a point source of harmonically pulsating strength
%\begin{equation}\label{eq:source}
%  Q(t) = Q_0 \cos (\omega t).
%\end{equation}
$Q(t) = Q_0 \cos (\omega t)$.
The source strength $Q(t)$ is the instantaneous volume flux
emitted from the singular source, which is located at the constant depth $z=-D$. We will go to complex notation for the time dependence henceforth. The mass balance is given by the continuity equation
\begin{equation}\label{eq:sourcepoint}
  %\hat{u}_x + \hat{v}_y + \hat{w}_z 
  \nabla\bdcdot\bv
  = Q_0 \delta(x) \delta(y) \delta(z+D) \rme^{-\rmi \omega t}.
\end{equation}
We shall seek solutions so the temporal behaviour of all perturbations is through a factor $\exp(-\rmi\omega t)$, i.e., the periodic steady state solution.

The variables are Fourier transformed as follows
\begin{equation}\label{eq:fourier}
  (\hat{u},\hat{v},\hat{w},\hat{p}) =  Q_0 \int \frac{\rmd^2 k}{(2\upi)^2}
  (u(z),v(z),w(z),p(z)) \rme^{\rmi \bd{k} \bdcdot \bd{r}_{\perp} - \rmi \omega t},
\end{equation}
where the real part has physical significance. We have introduced the horizontal position vector 
$\bd{r}_{\perp}=(x,y)=(r\cos\phi,r\sin\phi)$ and the wave vector $\bd{k} = (k_x,k_y)=(k \cos \theta, k \sin \theta)$, in Cartesian and polar phase space coordinates, respectively. 
Note that $k=|\bd{k}|$ which, unlike components $k_x,k_y$, is $\geq 0$, an important point to bear in mind when comparing to theory for the corresponding 2D flow in \ET\ where $k$ is used in the sense $k_x$. 
We let $\theta$ and $\phi$ take values between $-\upi$ and $\upi$. 

We shall use the terms upstream and downstream in the sense that they were used by \ET, by considering the source and undisturbed surface to be stationary. Thus we refer to directions $|\theta|<\upi/2$ as propagating upstream (ahead of the source), and $|\theta|> \upi/2$ are denoted downstream propagation directions. This would be opposite were we to consider the surface to be in motion, as was the case, e.g., in \citet{ellingsen14b}.

The transformed components of the Euler equation are given by
\bs
\begin{align}
- \rmi (\omega - k_x U) u + S w =& - \rmi k_x p/\rho,\label{eq:euler1}\\
- \rmi (\omega - k_x U) v =& - \rmi k_y p/\rho, \label{eq:euler2}\\
- \rmi (\omega - k_x U) w  =& - p'/\rho,\label{eq:euler3}
\end{align}
\es
where a prime denotes derivative with respect to $z$. 
The transformed continuity equation is
\begin{equation}\label{eq:continuity}
  \rmi k_x u + \rmi k_y v + w' = \delta (z+D).
\end{equation}
In exactly the same way, the surface elevation is Fourier transformed according to
\begin{equation}\label{eq:elevfourier}
  \zeta(x,y,t) = Q_0 \int \frac{\rmd^2 k}{(2\upi)^2} B(\bd{k})  \rme^{i \bd{k} \bdcdot \bd{r}_{\perp} - i \omega t}.
\end{equation}

{\color{red}
%%%%%%%%%%%%%%%% S E C T I O N %%%%%%%%%%%%%%%%%%
\subsection{Vorticity dynamics}\label{sec:vort}

Before continuing, let us take a moment to consider the vorticity equation for the system at hand, which to linear order in perturbations reads
\be\label{vort}
  \frac{\mathrm{D}\bOm}{\mathrm{D}t} = S\frac{\partial\bv}{\partial y} - S\bd{e}_y (\nabla\bdcdot\bv)=S\frac{\partial\bv}{\partial y} - SQ_0 \delta(x) \delta(y) \delta(z+D) \rme^{-\rmi \omega t}\bd{e}_y
\ee
where $\bOm=\nabla\times \bv$ is the perturbation vorticity, the background flow has vorticity $S\bd{e}_y$, and we have inserted \eqref{eq:sourcepoint}. The two terms on the right hand side correspond to two different mechanisms whereby the vorticity of a fluid particle might change as it travels with the flow. 

The term proportional to $\delta(x)\delta(y)\delta(z+D)$ was discussed at length in \ET. A fluid particle passing through the source point is given additional vorticity because its volume is momentarily changed while its circulation remains the same. The additional vorticity due to the inhomogeneity in the vorticity equation (the term proportional to $\delta(z=D)$) is expected to be proportional to $-SQ_0\exp(-\rmi\omega t)/U(-D)$, because the time such a small volume of fluid spends with the source inside of it is proportional to $1/U(-D)$. 

The term $S\partial\bv/\partial y$ comes from the fact that a wave propagating at an oblique angle over a shear current will gently shift and twist the vortex lines as it passes, as discussed by \cite{ellingsen16}. The effect may be understood in terms of the plane wave components of the wave picture \eqref{eq:elevfourier} by noting that $\partial/\partial y\to \rmi k_y$ in Fourier space. Thus it is the combination of the sub-surface shear $S$ and an oblique propagation direction relative to the current, $k_y\neq 0$, which causes the vorticity of fluid particles to change as they are convected.
}

%%%%%%%%%%%%%%%%%%%%%%%%%%%%%%%%%%%%%%%%%%%%%%%%%
%%%%%%%%%%%%%%%% S E C T I O N %%%%%%%%%%%%%%%%%%
%%%%%%%%%%%%%%%%%%%%%%%%%%%%%%%%%%%%%%%%%%%%%%%%%
\section{Formal solution of the radiation problem}\label{sec:solution}

From the transformed governing equations we can eliminate $u, v$ and $p$ to obtain an equation for the vertical velocity alone
\begin{equation}\label{eq:vertfourier}%Equation re-derived and checked 12.06.15
  w'' - k^2 w =  \delta'(z+D) - \frac{k_x S}{\omega - k_x U}\delta(z+D).
\end{equation}
Since $\delta(z+D)$ is nonzero only at $z=-D$, the fraction $k_x S/(\omega - k_x U)$ effectively becomes $k_x S/(\omega + k_x SD)$ in the last term.
The homogeneous solution of this equation which satisfies the bottom condition is written in the form
\begin{equation}\label{eq:homo}
  w_h(z) =  A(\bd{k}) \sinh k(z+H),
\end{equation}
where $A(\bd{k})$ is an undetermined velocity amplitude.
In addition, there are two particular solutions to \eqref{eq:vertfourier} due to the two forcing terms on the right hand side,
\begin{align}\label{eq:inhomo1}
  w_{p1}(z) =&  \cosh k(z + D) \Theta(z + D), \\
	\label{eq:inhomo2}
  w_{p2}(z) =& -\frac{k_x S}{k (\omega + k_x S D)} \sinh k(z +D) \Theta(z + D),
\end{align}
where the Heaviside unit step function $\Theta(z)$ has been introduced. The solution is strikingly similar to that obtained in the 2D case of \ET.

From transforming the kinematic free-surface condition (\ref{eq:surfkincond}) we find the relationship
\begin{equation}\label{eq:boga}
A \sinh kH = - \cosh kD + \frac{k_x S \sinh kD}{k(\omega + k_x S D)} -\rmi \omega B.
\end{equation}
Similarly, the dynamic free-surface condition (\ref{eq:dynamic}) gives
\begin{align}
  (\omega k \coth kH + k_x S)A\sinh kH  + \omega k \sinh (kD)& \notag \\
   + k_x S \cosh k D-\frac{\omega k_x S \cosh kD +(k_x^2 S^2/k) \sinh kD}{\omega + k_x S D} &= - \rmi g k^2 B.\label{eq:aogb}
\end{align}

From the two last equations we eliminate $A$, since we are most interested in the surface elevation. It is expressed by $B(\bd{k})$ 
so that
\begin{align}\label{eq:bformel}
  \frac{\rmi g}{\omega} \left(k - \frac{\omega}{g}  S \cos \theta  - \frac{\omega^2}{g} \coth kH \right) B(\bd{k}) = \frac{\cosh k(H-D)}{\sinh kH} + \frac{S\cos\theta\sin k(H-D)}{(\omega + kSD\cos\theta)\sinh kH}
\end{align}

%%%%%%%%%%%%%%%%%%%%%%%%%%%%%%%%%%%%%%%%%%%%%%%%%
%%%%%%%%%%%%%%%% S E C T I O N %%%%%%%%%%%%%%%%%%
%%%%%%%%%%%%%%%%%%%%%%%%%%%%%%%%%%%%%%%%%%%%%%%%%
\section{Dispersion relation and the radiation condition}\label{sec:disprel}

In \eqref{eq:bformel} $B(\bd{k})$ is multiplied by a factor that will be zero when 
$\bd{k}$
has a value corresponding to the dispersion relation for a given $\omega$,
\begin{equation}\label{eq:disprel}
  k - \frac{\omega}{g}  S \cos \theta  - \frac{\omega^2}{g} \coth k H = 0.
\end{equation}
This is identical to the dispersion relation derived and discussed at length by \citet{ellingsen14b}, except that here it is $\omega$, not $\bd{k}$, which is taken to be the known quantity. The dispersion relation was derived by \citet{tyvand14} for the cases $\theta=0$ and $\theta=\upi$ corresponding to 2D flow. See also the more detailed discussion of the 2D dispersion relation by \cite{ellingsen14c}. This selected value of the wave 
vector
gives a pole in the solution integral (\ref{eq:elevfourier}).

%%%%%%%%%%%%%%%%%% S E C T I O N %%%%%%%%%%%%%%%%%%%%%
\subsection{Dispersion relation in deep water}\label{sec:dispdeep}

Let us analyse the dispersion relation \eqref{eq:disprel}, which we treat as a function of $\bk$ with $\omega$ a fixed parameter. The following discussion generalises that found in Appendix A of \ET.
Wishing to consider the simpler case of infinitely deep water, $H\to \infty$, the naive procedure is to let $\coth kH\to 1$ in Eq.~(\ref{eq:disprel}), whereby the dispersion relation becomes simple and explicit with respect to $k$:
\be  \label{eq:kinf}
  k(\theta) = \frac{\omega}{g}\left(\omega + S\cos\theta\right).
\ee
We immediately see a potential problem with relation \eqref{eq:kinf}, because when $S>\omega$ it seems to permit negative values of $k(\theta)$, which is nonsensical and must be remedied by manually excluding the corresponding values of $\theta$ from the final integral. 

The reason for this conundrum may be found in the fact that mathematical limit taken in dispersion relation (\ref{eq:disprel}) is $kH\to \infty$, which is the same as the deep water limit $H\to\infty$ only if $k$ does not simultaneously tend to zero. A numerical solution $k(\theta)$ of dispersion relation (\ref{eq:disprel}) when $H$ grows large shows that in the ``forbidden sectors'' where \eqref{eq:kinf} is negative, the solution $k(\theta)$ tends quickly to zero so that $k(\theta)H$ remains finite near $\theta = \pm\upi$. In other words the forbidden sector is a phenomenon which only occurs at infinite depth, while for any finite $H$ a positive solution $k(\theta)$ exists in all directions. 

We can study this a little more closely by writing Eq.~\eqref{eq:disprel} on the form
\be\label{eq:etadisp}
  \coth \chi - \eta \chi = -(S/\omega)\cos\theta
\ee
where $\chi = k(\theta)H$ and $\eta = g/H\omega^2$. When the water is deep, $\eta\ll1$. The sectors in which \eqref{eq:kinf} predicts negative $K$ values are clearly where $\cos\theta<0$, near $\theta=\pm\upi$. Assuming therefore that the right hand side of \eqref{eq:etadisp} is positive and $>1$, there clearly exists a solution for moderate values of $\chi$ where the term containing $\eta$ can be neglected, approximated by 
\be\label{eq:KHasymp}
  k(\theta)H=\chi \buildrel{H\to\infty}\over{\approx} \mathrm{arcoth}(-S\cos\theta/\omega) = \frac12 \ln\frac{\omega-S\cos\theta}{-\omega-S\cos\theta}
\ee
provided $-\omega-S\cos\theta>0$.
Thus, when $S>\omega$ the product $k(\theta)H$ never grows large near $\theta=\pm \upi$ as $H\to \infty$, and the naive taking of the deep water limit leads to unphysical solutions. 

\begin{figure}
  \includegraphics[width=\textwidth]{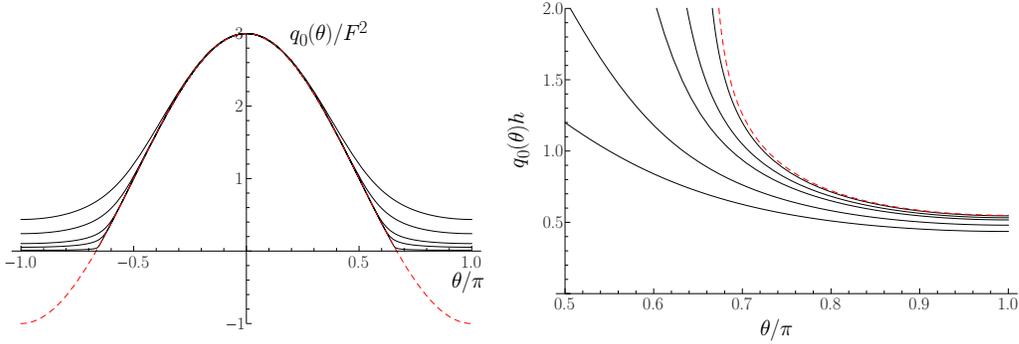}
  \caption{Left: Plots of wave numbers $k_0(\theta)$ which solve (\ref{eq:disprel}), non-dimensionalised as in Table \ref{tab:units}, for different values of $F^2 h = 1,2,5,10,50$ (solid graphs, top to bottom) with $S/\omega=2$. The naive expression \eqref{eq:kinf} is plotted as dashed line. Right: Plots of $q_0(\theta)h$ for the same solutions, plotted near $\theta=\upi$, again for $F^2 h = 1,2,5,10,50$ (solid graphs, bottom to top). The asymptotic solution \eqref{eq:KHasymp} is shown as dashed line.}
  \label{fig:K}
\end{figure}

Numerical solutions to the dispersion relation \eqref{eq:disprel} for the case $S=2\omega$ are shown in figure \ref{fig:K} for different values of the (suitably non-dimensionalized) depth $H$, where the relation \eqref{eq:kinf} is also shown. Said figure also shows the behaviour of $k(\theta)H$ near $\theta\to \upi$ when $H$ grows large. It is clear from the numerical solutions that the correct limiting dispersion relation is in fact not \eqref{eq:kinf}, but instead
\be
  k_0(\theta) \buildrel{H\to\infty}\over{=} \frac{\omega}{g}\left(\omega + S\cos\theta\right)\Theta(\omega + S\cos\theta)
\ee
with $\Theta$ again the Heaviside function. What this implies is that in propagation directions where $\omega + S\cos\theta<0$, the permissible plane wave solution approaches infinite wavelength. In a similar manner to that of the appendix of \ET\ one can show that there is no energy transport in this ``forbidden sector'' in the limit $H\to\infty$, and hence this sector is to be simply excluded from the integral over $\bk$ in the deep water limit by means of a cut-off: values of $\theta$ must satisfy
\be\label{eq:th0}
  |\theta| < \arccos\bigl[\max(-\omega/S,-1)\bigr] \equiv \theta_0.
\ee

For any $H<\infty$, the integral is not cut off explicitly. That said, when $S/\omega>1$ and $h\gg g/\omega^2$ (deep waters), the sector which at infite depth would be forbidden is characterised by regular waves of very long wavelength and correspondingly small amplitude.

%%%%%%%%%%%%%%%%%% S E C T I O N %%%%%%%%%%%%%%%%%%%%%
\subsection{Radiation condition}

\begin{table}
  \begin{center}
  \begin{tabular}{lll}
  Dimensional quantity& Non-dimensional quantity& Name \\
   $S$ & $\sigma=S/\omega$ & Shear parameter \\
   $H$ & $h=H/D$ & Relative depth\\
  $\zeta$ & $\tzeta=\zeta/\zeta_0$ & Nondimensional elevation \\
   $x,y,z$ & $(\tx,\ty,\tz)=(x,y,z)/D$ &Nondimensional positions\\
   $t$ & $T=t\omega$ &Phase\\
   $\bk$ & $\bq=D\bk$ & Wave vector \\ 
  \end{tabular}
  \caption{Table of non-dimensional quantities.}
  \label{tab:units}
  \end{center}
\end{table}

Before we proceed, let us nondimensionalise all quantities with respect to length $D$ and time $1/\omega$, particularly
\be
  \bq = D\bk, ~~\bx = \br_\perp/D,~~ T = \omega t,
\ee
and we obtain the dimensionless parameters
\begin{align}
  \sigma =& S/\omega, \\
  F =& \omega\sqrt{D/g}.
\end{align}
An exception from this nondimensional scaling is made for the perturbed quantities which are scaled instead with respect to the length scale
\be
  \zeta_0 = \frac{Q_0\omega}{gD}.
\ee
Here $\sigma$ is the dimensionless shear (or vorticity) and $F^2$ is a Froude number based on length $D$ and velocity $\omega D$. The various non-dimensional quantities are tabulated in Table~\ref{tab:units}. 
The expression for the surface elevation now becomes 
\bs\label{eq:zeta}
\begin{align}
  \tzeta =& \frac{ \rme^{-\rmi T}}{4\upi^2 \rmi}\int_{-\upi}^{\upi} \rmd \theta \int_0^\infty \rmd q \,q\frac{\rme^{\rmi q \tr\cos(\theta-\phi)}}{\Gamma(\bq)\sinh qh}\left[\cosh q(h-1) + \frac{\sinh q(h-1)}{q-q_c}\right], \\
  \Gamma(\bq) =& q - F^2(\coth qh + \sigma\cos\theta),
\end{align}
\es
where $\tzeta=\zeta/\zeta_0$, $h = H/D$, $\tr^2 = \tx^2+\ty^2$.%, and we defined the dimensionless amplitude factor $\alpha = Q_0\omega/gD^2$.
We also define the critical dimensionless wave number
\be
  q_c = -\frac1{\sigma\cos\theta}.
\ee

Inserting the solution for $B(\bk)$ in Eq.~\eqref{eq:bformel} into Eq.~\eqref{eq:elevfourier} now produces an integral which is not well defined because it contains poles on the integration axis.
The problem is a standard one encountered in all periodic or stationary wave systems, and must be resolved by means of a radiation condition ensuring that only outgoing waves are included while formal solutions to the strictly periodic problem which correspond to waves originating at infinity must be excluded. 
As in \ET\ we employ the radiation condition in the manner of, e.g., \citet[][\S4.9]{lighthill78} by presuming that the point source has been switched on slowly and gradually since $t=-\infty$, by replacing
\be
  Q_0 \to Q_0 \rme^{\epsilon\omega t}
\ee
where $\epsilon$ will be taken to zero in the end through positive values (this definition makes $\epsilon$ dimensionless, which is convenient). This amounts to a replacement rule everywhere,
\be\label{eq:replace}
  \omega \to \omega(1+\rmi\epsilon).
\ee

As $\epsilon$ is taken towards zero, its sole significance is that it moves the poles in the integrand slightly off the real $q$ axis. In exact analogy to the 2D case considered in \ET\ there are two types of poles in the integrand, corresponding to two types of waves. Firstly there are the standard dispersive waves at values $q_0(\theta)$ which satisfy $\Gamma(\bq)=0$, i.e., 
\be\label{eq:q0def}
  q_0 = F^2[\coth (q_0h) + \sigma\cos\theta]
\ee
which defines $q_0(\theta)$ implicitly. Secondly there is a pole at $q=q_c$ because of the presence of a 
{\color{black}
critical layer-like downstream flow, as discussed at length in \ET. 
}
It was shown in \ET\ how the critical layer in 2D is a sequence of vortices which is advected downstream with the flow at the velocity $U_c=-SD$ which the fluid has at depth $D$. The nature of the critical layer solution in the present 3D geometry will be analysed further below. 

The replacement rule \eqref{eq:replace} gives a revised expression for $\tzeta$ which is now uniquely defined:
\begin{align}\label{zfull}
    \tzeta =& \frac{ \rme^{-\rmi T}}{4\upi^2 \rmi}\int_{-\upi}^{\upi} \rmd \theta \lim_{\epsilon\to 0^+}F(\theta), \\
    F(\theta) =& \int_0^\infty \rmd q \,q\frac{\rme^{\rmi q \tr\cos(\theta-\phi)}}{[\Gamma(\bq)
    -\rmi\epsilon\Phi(\bq)]\sinh qh}\left[\cosh q(h-1) + \frac{\sinh q(h-1)}{q-q_c(1+\rmi\epsilon)}\right] ,\\
    \Phi(\bq) =& F^2(2\coth qh + \sigma\cos\theta).
\end{align}
This expression may now be used directly for numerical purposes by choosing a small but nonzero value of $\epsilon$. In the following we will also extract the far-field part of the solution in a manner similar to that of \ET.

%%%%%%%%%%%%%%%%%%%%%%%%%%%%%%%%%%%%%%%%%%%%%%%%%%%%%%
%%%%%%%%%%%%%%%%%% S E C T I O N %%%%%%%%%%%%%%%%%%%%%
%%%%%%%%%%%%%%%%%%%%%%%%%%%%%%%%%%%%%%%%%%%%%%%%%%%%%%
\section{Far-field solution}

As discussed extensively in the literature \citep[c.f., e.g.,][]{wehausen60,lighthill78}, it is the poles of the integral over $\bd{q}$, i.e., the poles of \eqref{eq:zeta}, which contribute to the far zone. This is true also in the present case as we argue below. Regard the integral as written in Eq.~\eqref{zfull}, so that the integral of the wave number (or wave vector modulus) $q$ is taken first, then the integral over propagation directions $\theta$ second. The physical significance of the pole near $q=q_0(\theta)$ is well known: $q=q_0$ is simply the dispersion relation which must be satisfied in order for waves to propagate towards infinity. 

The second pole, at $q=q_c(\theta)$ is physically interesting and was discussed for the 2D case in \ET. The pole at $q=q_c$ occurs for the wave number at which waves at depth $z=-D$ propagating in direction $\theta$  have a phase velocity which, when measured along the direction of the current, $-\bd{e}_x$, is equal to the velocity of the fluid flow, 
{\color{red}
$U(-D) = -SD$. 
}
This is the condition for the formation of a critical layer, a phenomenon which has been studied extensively in 2D. %Common wisdom holds that the critical layer is associated with a singularity in the Rayleigh equation \citep{booker67} which only occurs if the flow either has stratified density or $U''(z)\neq 0$. 
{\color{red}
We showed (\ET) that even for a linear shear current of uniform vorticity in 2D a critical layer-like flow is created by a submerged oscillating source. 
}
The existence of critical layer type solutions for the constant vorticity flow in 2D was previously reported in the mathematical literature \citep{ehrnstrom08,wahlen09}, and takes the form of a train of vortical fluid structures similar to Kelvin's cat's eye vortices, centered at the source depth and advected downstream with the shear flow. In 3D the same phenomenon occurs, and we will show that the surface deformation associated with the critical layer typically takes the form of a narrow train of waves which retains constant shape as it is convected downstream.

{\color{red}
Note that while we refer to the vortical flow drifting downstream of the source as a critical layer flow due to its obvious similarities with critical layers as they occur in the literature on waves on shear flows \citep[e.g.,][]{booker67}, there is an important difference regarding how the critical wave interacts with the propagating surface waves. In the present case, the critical layer is not caused by a propagating wave impinging on a critical level, but is created by the source itself, and does not interact with the propagating regular waves except near a particular resonance frequency, which we discuss in detail in the following.
}

\begin{figure}
  \includegraphics[width=\columnwidth]{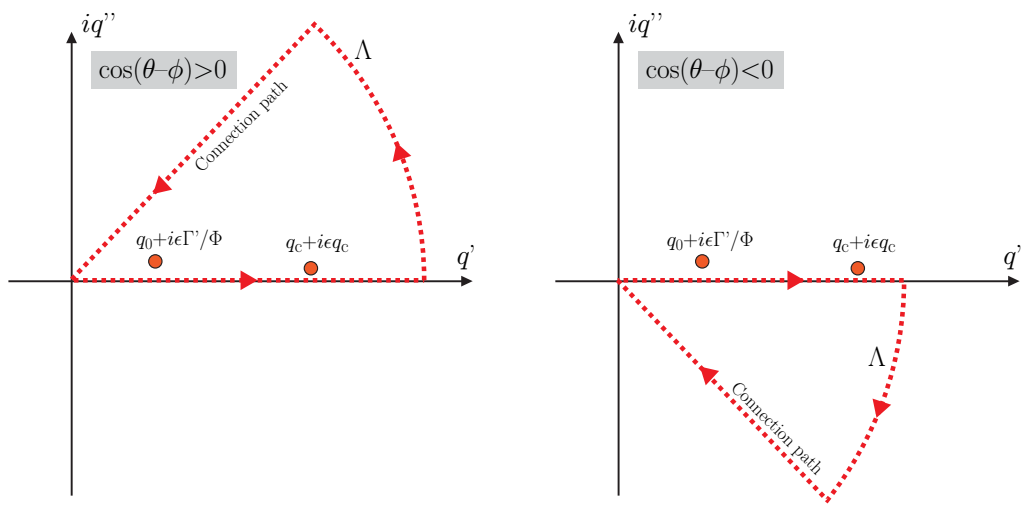}
  \caption{Contours in the complex $q$ plane used for extracting the far-field solution.}
  \label{fig:contours}
\end{figure}

In order to evaluate the far field we close the integration contour in the complex $q$ plane as shown in Fig.~\ref{fig:contours}. In order for the contribution from the poles to be identifiable as the far-field contribution it is necessary to close the contour above or below according to the sign of the exponent factor $\cos(\theta-\phi)$ to ensure that the integrand along each leg of the contour  is a decreasing function of $\tr$ as $\tr\to\infty$. We choose to close the contour by means of a circular arc at infinity and a straight path making an angle with the real axis which we may choose arbitrarily so long as it is strictly greater than zero (to ensure the above mensioned poles are enclosed when they should be), and small enough to exclude a series of poles a little to the side of the imaginary axis where $\Gamma(\bq)$ has zeros. These poles contribute to the near field only, and lie a little to the right or the left of the imaginary axis depending on $\cos\theta$ as well as $h,\sigma$ and $F^2$. For numerical purposes we choose a path making an angle $\gamma=\pm0.4\upi$ with the real axis, which caused no problems with the parameters we considered (it is numerically beneficial to be as close as possible to the path of steepest descent, which in the deep water case is at $\tan\gamma = \tr\cos(\theta-\phi)$, yet this path will lie perilously close to the imaginary axis for large $\tr$).

Consider now the two poles in the integrand of Eq.~\eqref{eq:zeta}. Upon replacing $\omega\to \omega (1+\rmi\epsilon)$ the poles have been shifted to the positions
\begin{align}
  q_c \to& q_c(1+\rmi\epsilon),\label{eq:qcC}\\
  q_0 \to& q_0 + \rmi\epsilon \Phi(q_0)/\Gamma'(q_0).\label{eq:qr}
\end{align}
Since the contribution from the pole to the integral $F(\theta)$ contributes the far field, it is clear that a far-field contribution from either regular or critical waves exists only for values of $\sigma, F^2$ and $\theta$ for which the pole lies inside the contour. A similar analysis is performed for the case of ship waves by \citet{li16b}.

Assume now that $q_0\neq q_c$ so that the two poles are well separated and may be treated independently (according to the terminology we will introduce in Sec.~\ref{sec:resonance}, a situation where this is true for all $\theta$ is called sub-resonant). We will need to determine whether they lie above or below the real axis.

The simplest pole to analyse is the critical one at $q=q_c$. From Eq.~\eqref{eq:qcC} it is clear that this pole can lie inside the contour only if $q_c>0$, i.e., $\cos\theta<0$, and in this case it will be shifted \emph{above} the complex axis, hence contribute to the far-field only if $\cos(\theta-\phi)>0$.

The propagating pole is a little more involved. For the pole to lie within the integration path it is necessary that a solution $q_0>0$ of \eqref{eq:q0def} exists, and that the pole is shifted to the appropriate side of the axis, either above or below depending on how the contour is closed. With
\be
  \Gamma'(q_0) = 1+\frac{F^2 h}{\sinh^2q_0h}
\ee
it is quite obvious that $\Gamma'(q_0)>0$ for all positive $q_0$, so the position of the pole \eqref{eq:qr} is decided by the sign of $\Phi(q_0)$. However, by using \eqref{eq:q0def} we may write
\be
  \Phi(q_0) = F^2\coth(q_0h) + q_0
\ee
which again is clearly a positive quantity provided $q_0>0$. Hence both poles lie above the complex axis provided $q_0,q_c>0$ and contribute to the far field provided $\cos(\theta-\phi)>0$.

As discussed in Sec.~\ref{sec:disprel}, a solution $q_0(\theta)$ exists for all $\theta$ and all values of $F^2$ and $\sigma$ so long as $h$ is finite, but $q_0(\theta)$ tends to zero in a sector $|\theta|>\theta_0$ when $h\to \infty$. Within this region no far-field propagating wave exists in this limit.

%%%%%%%%%%%%%%%%%% S E C T I O N %%%%%%%%%%%%%%%%%%%%%
\subsection{Resonance}\label{sec:resonance}

The 2D case considered in \ET\ uncovered that a resonance may occur between the regular downstream propagating wave and the critical wave (which only occurs downstream) if their wave numbers are the same. In the present 3D configuration we henceforth distinguish between a sub-resonant and a 
super-resonant regime. Physically, resonance occurs when the regular wave propagating in direction $\theta=\upi$ has the same phase velocity as the critical wave, i.e., the same velocity as the fluid current at depth $z=-D$. Note how this differs fundamentally from the Doppler resonance which may occur for a source which moves relative to the surface: in that case it is the group velocity which causes resonance when it coincides with the current velocity \citep[e.g.][]{wehausen60}.

The situation which we define as sub-resonant occurs for parameters $F^2$ and $\sigma$ which are so that $q_0(\theta)\neq q_c(\theta)$ for all $\theta$. In this case the two poles in the integrand of \eqref{eq:zeta} can be treated independently. In other words, in the sub-resonant regime the downstream regular wave and the critical layer wave have very different appearence and behaviour and their individual contributions to the far-field are easily identifiable. At a deeper level this is because the critical layer wave can never satisfy the dispersion relation $\Gamma(\bq)=0$ when $F^2$ is sub-resonant. 

\begin{figure}
  \begin{center}
  \includegraphics[width=\textwidth]{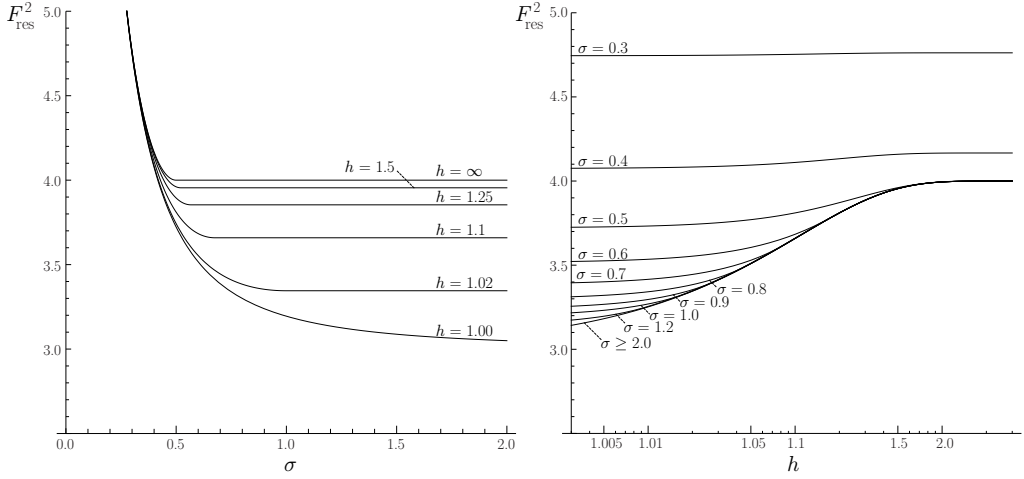}
  \end{center}
  \caption{$F_\text{res}^2$ as a function of $\sigma$ for different values of $h$ (left) and as a function of $h$ for different $\sigma$ (right). Any $h\gtrsim 2$ has $F_\text{res}$ virtually indistinguishable from infinite depth.}
  \label{fig:Fres}
\end{figure}

The criterion for the situation to be sub-resonant is that $\Gamma(q_c,\theta)\neq 0$ for all $\theta$. The equation $\Gamma(q_c,\theta)=0$ may be written
\be\label{Frescrit}
  F^2 = \frac1{\sigma\cos\theta\left(\coth\frac{h}{\sigma\cos\theta}-\sigma\cos\theta\right)}.
\ee
In sub-critical situations, this equation has no solution. The right hand side is either always greater than $F^2$ or, in some cases when $\sigma,h>1$, can be negative for some $\theta$. Since $F^2$ is positive by definition, its resonant value is the smallest value for which Eq.~\eqref{Frescrit} has a solution, corresponding to the smallest positive value of the right hand side of Eq.~\eqref{Frescrit}. 
These positive minima occur at $\theta=0,\pm\upi$ when $\sigma<0.5$, but above some $h$ dependent value of $\sigma$ the minima are found instead at two values of $\theta$ either side of $0$ and $ \upi$. After this bifurcation the minimum value of $F_\text{res}$ remains constant for higher values of $\sigma$. We define
\be\label{Fres}
  F_\text{res}^2 = \Min_{\theta,>0} \left\{ \frac1{\sigma\cos\theta\left(\coth\frac{h}{\sigma\cos\theta}-\sigma\cos\theta\right)}\right\}
\ee
but so that $F^2_\text{res}>0$. The notation means that the minimum value is found with respect to $\theta$, but so that that negative values are ignored. 

One may show that if $\coth(h/\sigma)<\sigma$, the right hand side of Eq.~\eqref{Frescrit} is negative in some sectors near $\theta=0$ and $\upi$; no resonance is then possible in these sectors. A necessary but insufficient criterion is that both $\sigma$ and $h$ exceed $1$. The phenomenon thus bears resemblance with deep water cut-off, discussed in Section \ref{sec:dispdeep}.

We term the situation sub-resonant for values of $F^2$ smaller than $F^2_\text{res}$. In the super-resonant regime, $F^2>F^2_\text{res}$, and there always exists a value of $\theta$ for which the critical wave satisfies the dispersion relation. At resonance, the critical wave satisfies the dispersion relation at one or two distinct values of $\theta$ corresponding to the minimum or minima in equation \eqref{Fres}. The behaviour of $F^2_\text{res}$ for different $\sigma$ and $h$ is shown in figure \ref{fig:Fres}. For small $\sigma$, the critical wave satisfies the dispersion relation at $\theta=\upi$, and $F^2$ is a decreasing function of $\sigma$. For $\sigma$ above some threshold value, however, the critical wave first satisfies the dispersion relation (i.e., the minimum of the function in Eq.~\eqref{Fres}) for a pair of propagation angles either side of $\theta=\upi$. From this point on, $F^2_\text{res}$ remains constant as $\sigma$ is further increased.

The criterion for resonance is similar to that found in the 2D case, but the interpretation is now a little different: In 2D, resonance occured only at the single value $F^2=F^2_\text{res}$. At this value, waves continued to grow indefinitely in amplitude away from the source as a linear function of difference, while at values above or below $F^2_\text{res}$ wave amplitude remained bounded everywhere. A study of the behaviour of $F^2_\text{res}$ as a function of $h$ and $\sigma$ was given in \ET\ and needs not be repeated here. Here the situation is different mathematically in that a confluence of the two poles will occur at some $\theta$ for \emph{every} value $F^2>F^2_\text{res}$, although we shall see that also in the 3D case unbounded wave amplitudes can occur only exactly at the resonant value. Like in \ET, thus, the physical wave field only resonates at a single value $F^2 = F^2_\text{res}(\sigma,h)$.

%%%%%%%%%%%%%%%%%% S E C T I O N %%%%%%%%%%%%%%%%%%%%%
\subsection{Sub-resonant case}\label{sec:farfield}

\begin{figure}
  \includegraphics[width=\textwidth]{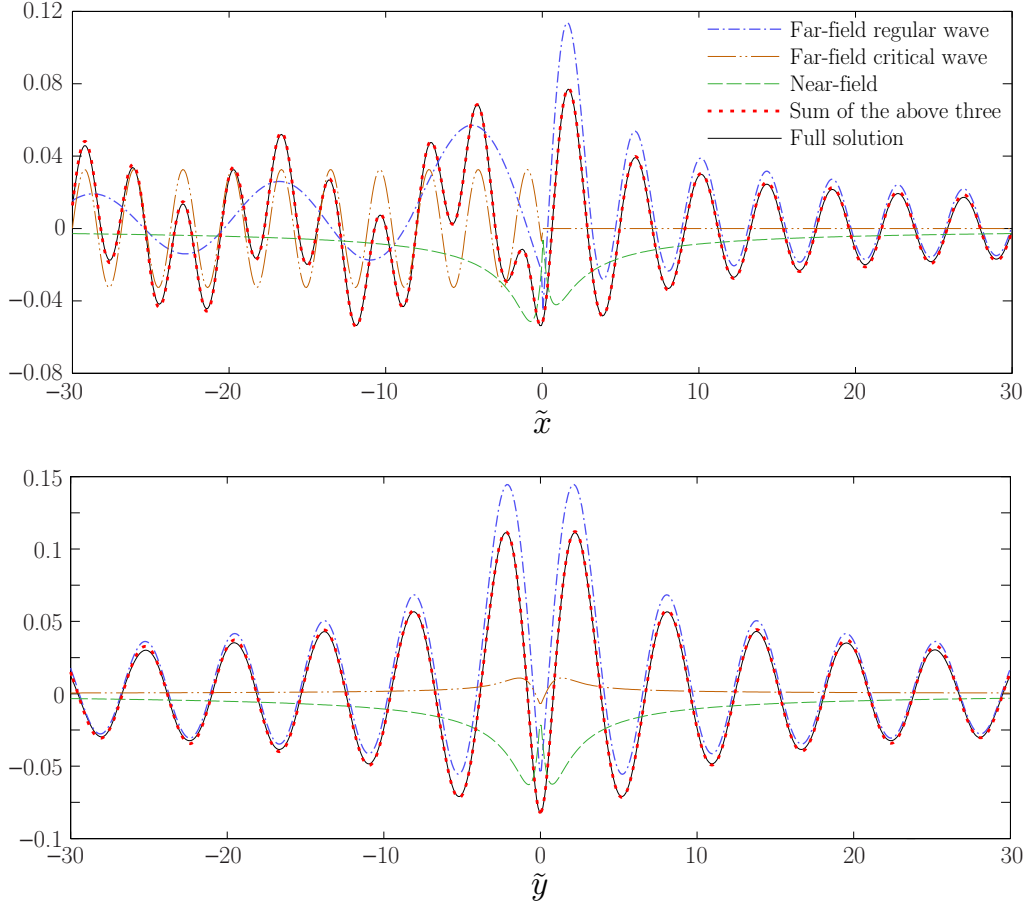} 
  \caption{Plots along the $\tx$ axis at $\ty=0$ (top) and along the $\ty$ axis at $\tx=0$ (bottom) of the full surface elevation calculated from Eq.~\eqref{zfull} (solid line) as well as numerical calculation of the far fields from regular as well as critical waves, and numerical calculation of the near-field from the connection integration path shown in Fig.~\ref{fig:contours}. Parameters in this example are $F^2=1, \sigma=0.5, h=4, T=0.3$.}
  \label{fig:nearfar}
\end{figure}

Assume now that $F^2<F^2_\text{res}$ so that $q_0\neq q_c$ for all $\theta$.
With reference to Eq.~\eqref{eq:zeta}, let's define
\be
  F(\theta) = \int_0^\infty\rmd q f(q,\theta)
\ee
and note that
\bs
\begin{align}
  \Res_{q=q_0}f(q,\theta)=& \frac{\rme^{\rmi q_0\tr\cos(\theta-\phi)}}{\Gamma'(q_0)\sinh q_0 h}\left[\cosh q_0(h-1) + \frac{\sinh q_0(h-1)}{q_0-q_c}\right] \\
  \Res_{q=q_c}f(q,\theta)=& \frac{\rme^{\rmi q_c\tr\cos(\theta-\phi)}}{\Gamma(q_c)}\frac{\sinh q_c(h-1)}{\sinh q_c h} 
\end{align}
\es
We bear in mind that the polar angle $\theta$ is the angle in the $\bq$ Fourier plane, and $\phi$ is the angle in the $\tx,\ty$ plane.
The integral around the closed contour $\Lambda$ gets no contribution from the circular arc at infinity, and is now
\begin{align}
  \oint_\Lambda \rmd q f(q,\theta) =& F(\theta) + F_\text{conn.}(\theta) \notag\\
  =& 2\upi \rmi \left[\Res_{q=q_0}f(q,\theta)+\Theta(-\cos\theta)\Res_{q=q_c}f(q,\theta)\right]\Theta[\cos(\theta-\phi)].
\end{align}
Here $F_\text{conn.}(\theta)$ is the integral along the connection path as shown in the figure, directed towards the origin. 

We ascertain numerically that the contribution from the connection path falls off faster than the far-field as $\tr\to\infty$ (it is possible, but cumbersome, to show this analytically using asymptotic methods). Hence we shall identify the contribution from the poles as the far field, while noting that the division of $\zeta$ into near field and far field is not a unique operation. 

The far-field surface elevation in the sub-resonant case is thus
\begin{align}
  \zff =& \frac{ \rme^{-\rmi T}}{2\upi}\int_{\phi-\upi/2}^{\phi+\upi/2}\rmd \theta \frac{q_0\rme^{\rmi q_0\tr\cos(\theta-\phi)}}{\Gamma'(q_0)\sinh q_0 h}\left[\cosh q_0(h-1) + \frac{\sinh q_0(h-1)}{q_0-q_c}\right] \notag \\
  &+ \frac{ \rme^{-\rmi T-\rmi\tx/\sigma}}{2\upi}\int_{\upi/2}^{3\upi/2}\rmd\theta \frac{q_c\rme^{-\rmi (\ty/\sigma)\tan\theta}}{\Gamma(q_c)}\frac{\sinh q_c(h-1)}{\sinh q_c h} \Theta[\cos(\theta-\phi)] \label{zff}
\end{align}
where we have inserted the definition of $q_c$ in the exponent of the corresponding integral to highlight that the $\tx$-dependence is of a pure plane wave form.

In Fig.~\ref{fig:nearfar} we moreover show an example numerical calculations of the full surface elevation from Eq.~\eqref{zfull} as well as the asymptotic regular and critical waves and the numerical evaluation of the connection path integral which we identify as the near field. The graph shows with clarity that the sum of the three contributions exactly matches the full integral as they should, a useful validation of the numerical code. 

\begin{figure}
  \includegraphics[width=\textwidth]{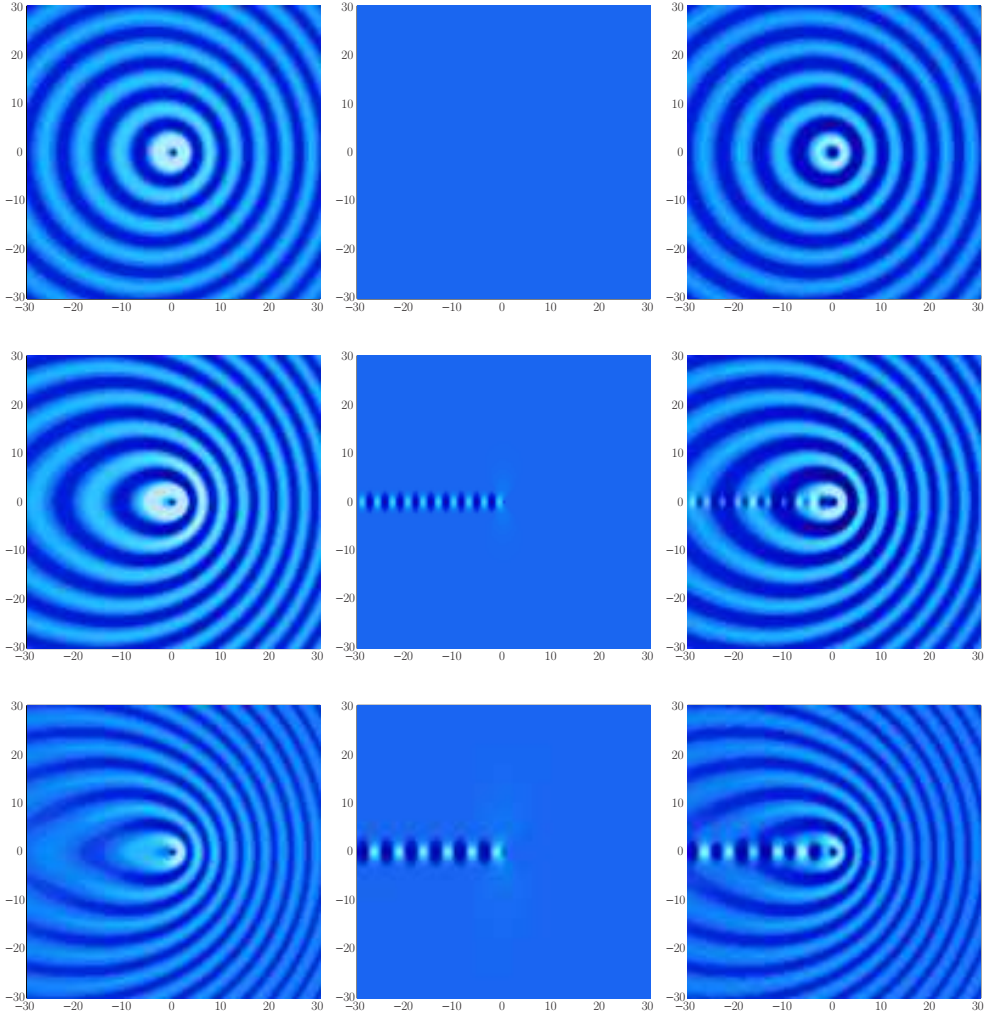} 
  \caption{Relief plots of typical sub-resonant wave field in the $\tx,\ty$ plane from oscillating source with $F^2=1, h=4$ and (top to bottom) $\sigma = 0.2, 0.5$ and $0.8$ at time $T=0.3$. Left: far-field solution for only the regular wave. Centre: far-field solution for only the critical wave. Right: full solution calculated from Eq.~\eqref{zfull} with $\epsilon=0.001$. Colour scaling is the same within each row. For the topmost row, the amplitude of the critical wave is so small as to be invisible.}
  \label{fig:subres}
\end{figure}

Let us study for a moment the last integral in Eq.~\eqref{zff}, which gives the critical ``wave''. When $\tr$ grows large, the exponential function typically becomes rapidly oscillating, and the integral is a rapidly decreasing function of $\tr$ according to the Riemann-Lebesgue lemma \citep[][\S6.5]{bender91}. This is true everywhere in the $\tr,\phi$ plane except along the line $\phi=\upi$ (where $\ty=0$) where the exponent becomes independent of $\theta$ (this is also true at $\phi=0$, but this line is excluded from the integral by the Heaviside $\Theta$ function). What this implies is that the critical wave must die quickly away from the source in all directions except along the line $\phi=\upi$ where it takes the form
\be
  \tzeta_{\text{f.f.},c}(\tx,\ty=0)=\frac{\rme^{-\rmi T-\rmi\tx/\sigma}}{2\upi}\Theta(-\tx)\int_{\upi/2}^{3\upi/2}\rmd \theta\frac{q_c }{\Gamma(q_c)}\frac{\sinh q_c(h-1)}{\sinh q_ch}.
\ee
The integral is now just a constant 
{\color{red}
with respect to $x$, 
}
and the behaviour along the ray $\phi=\upi$ is a wave of dimensionless wavelength $\lambda_c = 2\upi\sigma$, as found for the critical layer wave also in the 2D case by \ET;
{\color{red}
$\lambda_cD$ is the distance travelled during one oscillation period by a fluid particle at depth $z=-D$.
}

The deep-water limit of $\zff$ is 
\begin{align}
  \zff \buildrel{h\to \infty}\over{\longrightarrow}& \frac{ \rme^{-\rmi T}}{2\upi}\int_{-\theta_0}^{\theta_0}\rmd \theta q_0 \rme^{\rmi q_0\tr\cos(\theta-\phi)-q_0}\left(1 + \frac{1}{q_0-q_c}\right)\Theta[\cos(\theta-\phi)] \notag \\
  &+ \frac{ \rme^{-\rmi T-\rmi\tx/\sigma}}{2\upi}\int_{\upi/2}^{3\upi/2}\rmd\theta \frac{q_c\rme^{\rmi(\ty/\sigma)\tan\theta - q_c}}{q_c - F^2(1+\sigma\cos\theta)}\Theta[\cos(\theta-\phi)] ,
\end{align}
with $\theta_0$ defined in Eq.~\eqref{eq:th0}.

Numerical examples of sub-resonant wave fields are shown in Fig.~\ref{fig:subres} for increasing values of $\sigma$. The asymptotic far-field expressions are reasonable approximations for distances more than a couple of wavelengths from the source, and are numerically far quicker and also easier to calculate. 
%(for example, the surface calculations for the plots in Fig.~\ref{fig:subres} with $121\times121$ points take about 1 minute each on a standard desktop computer for the far-field expressions, and more than $10$ minutes for the double integral Eq.~\eqref{zfull} even after the latter calculation had been optimised for speed). 
Numerical evaluation of the full expression in Eq.~\eqref{zfull} is considerably more subtle than the corresponding calculation in the 2D case, because it was necessary to use a smaller value of $\epsilon$ when going to $\tr$ as large as $30$, giving very sharp contributions near the poles making brute force methods based on adaptive grid refinement very slow to converge. The chosen method was to calculate the position of the poles before integrating then pre-calculating an integration grid with much finer resolution near poles, whereupon integration was performed without further grid refinement. This method was able to speed up the integration by two orders of magnitude, at the expense of not knowing the exact integration accuracy, which was no problem in our case since both the far-field  and near-field solutions could be calculated arbitrarily accurately, providing a benchmark and convergence check such as shown in Fig.~\ref{fig:nearfar}.

\begin{figure}
  \includegraphics[width=\textwidth]{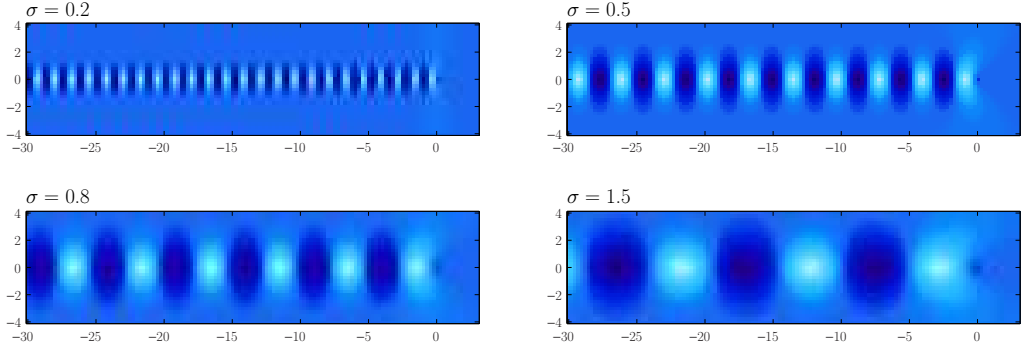}
  \caption{Relief plots of sub-resonant critical wave in the $\tx,\ty$ plane for increasing values of $\sigma$ (top left to bottom right) $\sigma=0.2, 0.5, 0.8, 1.5$. In all graphs: $F^2=1, h=4$. The colours are scaled individually in each panel and do not reflect relative amplitudes, which depend strongly on $\sigma$.}
  \label{fig:critwave}
\end{figure}

%%%%%%%%%%%%%%%%%% S E C T I O N %%%%%%%%%%%%%%%%%%%%%
\subsection{Deep or shallow water?}

The terms ``deep'' or ``shallow'' water in relation to surface waves are determined by the ratio of wavelength to water depth. While our dimensionless units were found to produce the best set of parameters ($F^2,h,\sigma$) to characterise the wave field in different situations, it somewhat disguises the effect of water depth since $h$ is a measure of how close the source is to surface or bottom, and bears no relation to the wavelength. 

Our sub-resonant far-field considerations reveal, however, that no single parameter on its own can hope to determine whether the system is ``shallow'' or ``deep'', simply because the critical waves and regular waves can have very different wavelengths. The criterion for a wave $\bq$ to be of ``deep water'' kind is that $qh\gg1$. Now, as we have seen, regular waves in deep water have wave numbers approximately $F^2(1+\sigma\cos\theta)$ (but see the discussion of `cut-off' above, for the case when $1+\sigma\cos\theta<0$), hence the criterion for regular waves to be ``deep'', at least in some directions,
is that $F^2(1+\sigma)h\gg 1$. On the other hand, the dimensionless wave number of the critical wave is $1/\sigma$ directly downstream, so the criterion that the critical wave is ``deep water'' is that $h\gg \sigma$. Often, the critical wave will be ``deep water'' in the sub-resonant regime even though the regular wave might be affected by the presence of finite depth.% Note that the transition to deep water is exponential so that $``\gg''$ in practice means ``more than a factor $2$ greater'' or similar, depending on the accuracy required. 

%%%%%%%%%%%%%%%%%% S E C T I O N %%%%%%%%%%%%%%%%%%%%%
\subsection{Asymptotic critical wave}

A case which is particularly suited for asymptotic analysis is when the shear is relatively weak or the frequency is high, so that we may assume $\sigma\ll1$. With this assumption the factor $\exp(-1/\sigma\cos\theta)$ in the integrand of the critical wave term of Eq.~\eqref{zff} means that only angles $\theta$ close to $\theta=\upi$ will contribute since the exponential is otherwise very small. We let $\beta = \theta-\upi$ and approximate the integral using Laplace's method \citep[][\S6.4]{bender91} by expanding the exponent around $\beta=0$ and taking $\beta=0$ elsewhere. Using
\begin{align*}
  q_c\cos(\theta-\phi) \sim -\frac1{\sigma}(\cos\phi + \beta\sin\phi + \mathcal{O}(\beta^3)); ~~ q_c \sim \frac1\sigma (1+ \half\beta^2+ \mathcal{O}(\beta^4)),
\end{align*}
the critical layer term is approximately 
\begin{align}\label{zffc}
  \tzeta_{\text{f.f.},c}(\tx,\ty)\sim& \frac{(1/\sigma)\rme^{-\rmi T-\rmi\tx/\sigma}}{2\upi\Gamma(1/\sigma,\upi)}\frac{\sinh\frac{h-1}{\sigma}}{\sinh \frac h\sigma}\Theta(-\tx)\int_{-\infty}^\infty\rmd \beta \rme^{-\rmi(\ty/\sigma)\beta-\beta^2/2\sigma}\notag \\
  \sim& \frac{\sinh\frac{h-1}{\sigma}}{\Gamma(1/\sigma,\upi)\sinh \frac h\sigma}\Theta(-\tx) \rme^{-\rmi T-\rmi\tx/\sigma} \frac1{\sqrt{2\upi\sigma}}\rme^{-\ty^2/2\sigma}
\end{align}
when $\sigma\ll 1$, where
\be
  \Gamma(1/\sigma,\upi) = \frac1\sigma - F^2\Bigl(\coth\frac h\sigma - \sigma\Bigr).
\ee 
We see that the surface elevation from the critical layer is restricted to a thin area close to the $\tx$ axis downstream of the wave of constant thickness $\sim\sigma$, falling off away from the axis in a Gaussian manner. In the absence of dissipation it does not decay away from the source, as is required by energy conservation in linear theory since the critical wave does not spread but travels downstream atop a row of vortices.

\begin{figure}
  \begin{center}
    \includegraphics[width=.7\textwidth]{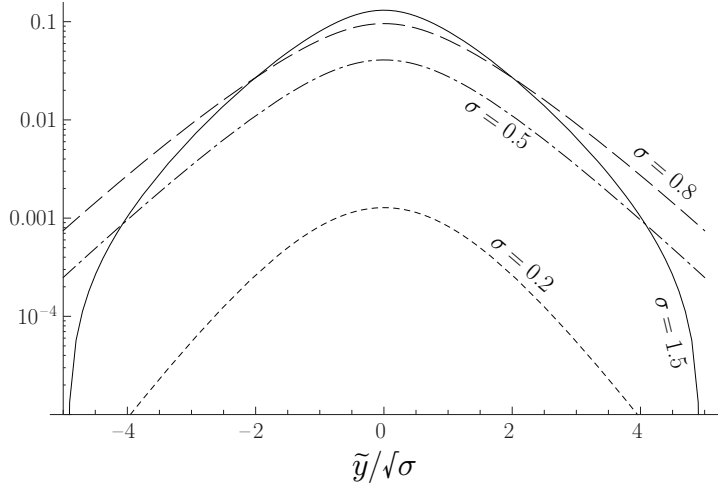}
  \end{center}
  \caption{Asymptotic amplitude function $f(\ty)$ for the critical waves shown in Fig.~\ref{fig:critwave}.}
  \label{fig:critamp}
\end{figure}

Although we assumed $\sigma$ small, Eq.~\eqref{zffc} is in fact a fair representation of the critical wave in sub-resonant cases for $\sigma\lesssim 0.4$. In any case the critical wave always has simple form $\tzeta_{\text{f.f.},c} \sim f(\ty) \exp(-\rmi T-\rmi\tx/\sigma)$ where the function $f(\ty)$ is an integral that is simple to evaluate numerically,
\be
  f(\ty) = \frac1{2\upi}\int_{\upi/2}^{3\upi/2}\rmd\theta\frac{q_c\cos[(\ty/\sigma)\tan\theta]}{\Gamma(q_c,\theta)}\frac{\sinh q_c(h-1)}{\sinh q_c h}
\ee
when $|\tx|\gg\ty$ (i.e., $\phi\approx \upi$). The amplitude function $f(\ty)$ for the four cases shown in Fig.~\ref{fig:critwave} are shown in Fig.~\ref{fig:critamp}. The distribution has approximately Gaussian shape with half-width $\sim\sqrt{\sigma}$ when $\sigma\lesssim 1$, but becomes distinctly non-Gaussian when $\sigma >1$.

The phase velocity of the critical wave in dimensional units is easily found to be $SD$, meaning that the critical ``wave'' is in fact the surface manifestation of an underwater row of flow structures which are convected with the mean flow, in the same way as found in the 2D case. Also similar to the 2D case, the amplitude of the critical wave diverges when $\Gamma(1/\sigma)\to 0$ which is the resonance criterion of Eq.~\eqref{Fres}.

%%%%%%%%%%%%%%%%%% S E C T I O N %%%%%%%%%%%%%%%%%%%%%
\subsection{Asymptotic regular wave}

The regular wave far-field expression, the first term of Eq.~\eqref{zff}, may be approximated further for large $\tr$ by use of the stationary phase approximation. At large $\tr$ the integrand genearally oscillates rapidly as $\exp(\rmi\tr \varphi(\theta))$ where the exponent function is
\be
  \varphi(\theta) = q_0(\theta)\cos(\theta-\phi).
\ee
At large distances the integral will be dominated by the value of $\theta$ where $\varphi'(\theta)=0$, i.e.,  a stationary point. Let the stationary point be called $\theta_s$. It is not possible to find explicit solutions for $\theta_s$, but it may be defined implicitly and solved numerically. 

Using the stationary phase approximation \citep{bender91} the regular far-field integral becomes, to leading order in $1/\tr$,
\be\label{zffr}
  \tzeta_{\text{f.f.},r}(\tr,\phi)\sim\frac{q_{0s}\rme^{\rmi\tr\varphi(\theta_s)-\rmi T + \sign[\varphi''(\theta_s)]\upi/4}}{\Gamma'(q_{0s})\sinh q_{0s}h\sqrt{2\upi\tr |\varphi''(\theta_s)|}}\left[\cosh q_{0s}(h-1)+\frac{\sinh q_{0s}(h-1)}{q_{0s}-q_c(\theta_s)}\right]
\ee
where we use shorthand $q_{0s}=q_0(\theta_s)$. 

It is worth noting before proceeding how the critical wave affects the amplitude of the regular wave indirectly, through the last term in the brackets in \eqref{zffr}. In the 2D case (\ET) the corresponding term was responsible for the regular wave being different in amplitude from that predicted by potential theory, sometimes radically so. In 3D no potential theory exists of course, but at least in the sub-resonant regime it is still meaningful to consider this term as the amplitude correction to the regular wave stemming from the fact that part of the mass flux from the source must go into the critical wave.

To calculate this asymptotic wave, we must find $\theta_s$ and $q_{0s}$ by solving a combined system of equations: The dispersion relation 
{\color{black}
\eqref{eq:q0def} 
}
and $\varphi'(\theta_s)=0$ which read, respectively,
{\color{black}
\be
   \Gamma(q_{0s},\theta_s)=0~~~\text{and} ~~~
   q_0'(\theta_s)\cos(\theta_s-\phi)-q_0(\theta_s)\sin(\theta_s-\phi)=0
   %\frac{\sigma F^2 \sin\theta_s\cos(\theta_s-\phi)}{\Gamma'(q_{0s})}+q_{0s}\sin(\theta_s-\phi)=0 
\ee
}
(note that $\Gamma(q,\theta)$ is a function of both $q$ and $\theta$ while $\Gamma'(q)$ is a function of $q$ only). We require the following relations, obtained by noting that $q_0$ satisfies $\Gamma(q_0)=0$,
\bs
\begin{align}
  \Gamma'(q_0) =& 1+hF^2[\coth^2q_0 h-1], \\
  \Gamma''(q_0) =& -2h^2F^2\coth q_0 h[\coth^2q_0 h-1],\\
  \coth q_0 h=& q_0/F^2 - \sigma\cos\theta, \\
  \varphi''(\theta)=& q_0''(\theta) \cos(\theta-\phi)-2q_0'(\theta)\sin(\theta-\phi)-q_0(\theta)\cos(\theta-\phi),\\
  q_0'(\theta) =& -\sigma F^2 \sin\theta/\Gamma'(q_0), \\
  q_0''(\theta) =& -\sigma F^2\cos\theta/\Gamma'(q_0)-(\sigma F^2\sin\theta)^2\Gamma''(q_0)/[\Gamma'(q_0)]^3.
\end{align}
\es

In deep water the calculation is significantly simpler; the asymptotic regular wave now reads
\be\label{eq:zregasympdeep}
    \tzeta_{\text{f.f.},r}(\tr,\phi)\sim\frac{q_{0s}\rme^{\rmi\tr \varphi(\theta_s)-\rmi T + \sign[\varphi''(\theta_s)]\upi/4}}{\sqrt{2\upi\tr |\varphi''(q_{0s})|}}\rme^{-q_{0s}}\left[1+\frac1{q_{0s}-q_c(\theta_s)}\right]
\ee
with $q_{0s}=F^2(1+\sigma\cos\theta_s)$ and where $\theta_s$ solves
\be%[checked 26.02.16]
  \sigma \sin(2\theta_s-\phi) + \sin(\theta_s-\phi)=0.
\ee
In fact, an explicit solution to this equation exists, but it is sufficiently complicated that numerical evaluation was found to be preferable.
Also the evaluation of $\varphi''(q_0)$ is now simple, with $q_0'(\theta) = -\sigma F^2\sin\theta$ and $q_0''(\theta) = -\sigma F^2\cos\theta$. 

The deep water stationary phase approximation is not valid near the line $\phi=\upi$, where $\varphi''(\theta_s)\to 0$. Exactly on this line, thus, the wave pattern falls off as $\tr^{-1/3}$ instead of as $\tr^{-1/2}$. Also for finite water depth the predicted amplitude very close to $\phi=\upi$ tends to be overestimated by Eq.~\eqref{zffr} (which is otherwise excellent), and more so when the deep water limit ($F^2 h\gg 1$ in this case) is approached.

%%%%%%%%%%%%%%%%%% S E C T I O N %%%%%%%%%%%%%%%%%%%%%
\subsection{Resonant case}

\begin{figure}
  \includegraphics[width=\textwidth]{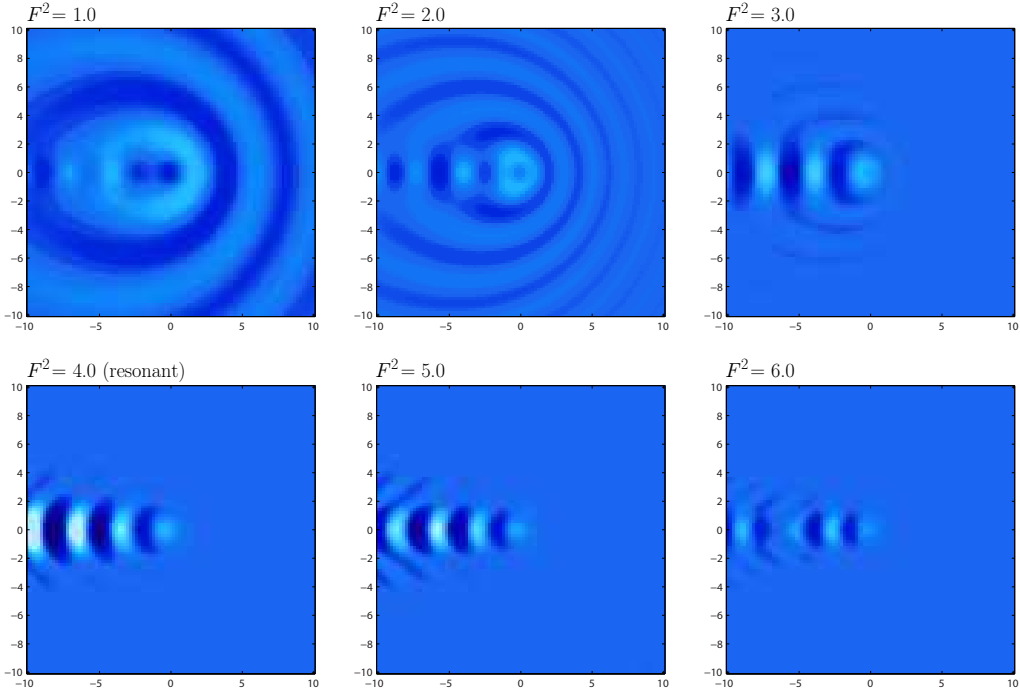}
  \caption{Relief plots of wave field $\tzeta(\tx,\ty)$ for increasing values of $F^2$ from sub-resonant to super-resonant (top left to bottom right) $F^2=1,2,3,4,5,6$. In all graphs: $\sigma=0.5, h=4, \epsilon=0.01$ and $T=0.3$. The colour range is from $-0.2$ (black) to $0.2$ (white).}
  \label{fig:incF}
\end{figure}

\begin{figure}
  \includegraphics[width=\textwidth]{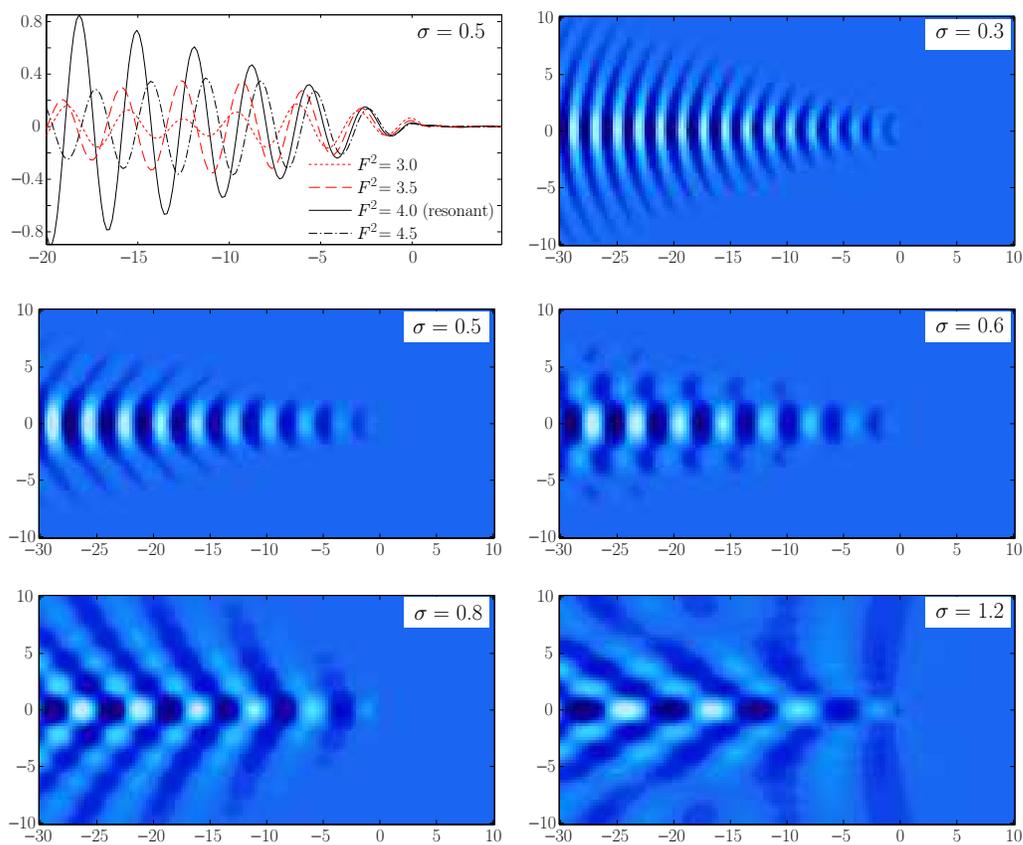}
  \caption{Upper left panel: Plots of $\tzeta(\tx,\ty=0)$ showing the transition from sub- to super-resonant for $\sigma=0.5, h=4$. Remaining 5 panels: Relief plots which show $\tzeta_\mathrm{f.f.}(\tx,\ty)$ for $F^2$ close to the resonant value (we choose $F^2 = F^2_\text{res}-0.1$) to illustrate the behaviour at resonance. According to Eq.~\eqref{Fres} with $h=4$, the Froude number is $F^2=4.66$ for $\sigma=0.3$, and $F^2=3.9$ for the remaining relief plots. $T=0.3$ everywhere.}
  \label{fig:resonant}
\end{figure}

The qualitative behaviour of the waves as $F^2$ is brought from sub-resonant to super-resonant may be seen from Fig.~\ref{fig:incF}, here for a moderate value of the shear, $\sigma=0.5$ and essentially deep water, $h=4$. The value of $F^2_\text{res}$ is $ 3.9999982$, so the case $F^2=4.0$ illustrates the resonating frequency. At low values of $F^2$, the regular propagating waves dominate, with the critical wave only a minor correction. As $F^2$ increases, the regular wave rapidly disappears, as may understood by regarding the asymptotic deep water far-field expression, Eq.~\eqref{eq:zregasympdeep}, where we see the amplitude of the regular waves decreasing as $\exp(-q_0)$. Since $q_0\propto F^2$ in deep water, it is clear that the critical wave must dominate as $F^2$ grows larger, because $q_c$ is independent of $F^2$ and no corresponding exponential decrease is found for the critical wave. Near $\phi=\upi$, however, the regular waves do not vanish, because regular wave of wavelength close to the critical wave's wavelength are amplified. Within a wedge around $\phi=\upi$ also the regular waves continue to grow when resonance is approached, interfering strongly with the critical wave. At $\phi=\upi$, thus, the behaviour is similar to that obtained in 2D; the regular and critical waves each diverge in amplitude as $F^2\to F^2_\text{res}$, but the combined wave increases linearly in amplitude as a function of $|x|$ downstream. This is shown in Fig.~\ref{fig:resonant}. 

The wave pattern at (or close to) resonance is only found downstream. Its appearence is that of the critical wave, with regular wave ``fringes''. The combined wave near resonance moreover increases linearly in width downstream, covering a constant angle. At the centre of the wedge is found the typical critical wave pattern, with regular waves outside it producing a herring bone pattern. The waves are contained within a wedge angle which depends on $\sigma$.

When the Froude number increases beyond $F^2_\text{res}$, wave amplitudes no longer grow indefinitely in the downstream direction, but are modulated in the manner typical of interfering waves with similar but unequal wavelength. The pattern remains within a downstream wedge, and distinct contributions from regular and critical waves once again become visible, although the formal splitting into two-component far-field is no longer possible in the manner of Sec.~\ref{sec:farfield}.

%%%%%%%%%%%%%%%%%%%%%%%%%%%%%%%%%%%%%%%%%%%%%%%%%%%%%%
%%%%%%%%%%%%%%%%%% S E C T I O N %%%%%%%%%%%%%%%%%%%%%
%%%%%%%%%%%%%%%%%%%%%%%%%%%%%%%%%%%%%%%%%%%%%%%%%%%%%%
\section{Some notes on the velocity field}

We provide here the formulas for the velocity field, although a detailed analysis of the velocity field as a function of coordinates as well as parameters $\sigma,F^2$ and $h$ is a major undertaking which we largely leave for a later investigation.

We found the vertical velocity component in section \ref{sec:solution} as
\be
  w(\tz) = A\sinh q(\tz+h) + \Bigl[\cosh q(\tz+1)-\frac{\sinh q(\tz+1)}{q-q_c}\Bigr]\Theta(\tz+1)
\ee
with
\be
   A(\bq) = \frac{1}{\Gamma(\bq)}\left[\left(F^2+\frac{q-\sigma F^2\cos\theta}{q-q_c}\right)\frac{\sinh q}{\sinh qh}-\left(q-\sigma F^2\cos\theta +\frac{F^2}{q-q_c}\right)\frac{\cosh q}{\sinh qh} \right]
\ee
with $\Gamma(\bq)=q-F^2(\coth qh + \sigma \cos\theta)$ as before. Note correspondence with Section 6 of \ET. From eliminating the pressure from Eqs.~\eqref{eq:euler2} and Eqs.~\eqref{eq:euler3} gives
\be
{\color{red}
  v'(\tz) - \frac{\sigma q_x}{1-\sigma q_x \tz}v(\tz) = \rmi q_y w(\tz).
  }
\ee
The homogeneous solution of this equation,
\be
  v_\text{homog.} = \frac{C}{1-\sigma q_x \tz}
\ee
can only agree with the dynamic boundary condition \eqref{eq:surfcond2} if $C=0$. The solution is thus the particular solution only, given by
\begin{align}
  v(\tz)=& \frac{\rmi q_y}{q}\left( A(\bq)\cosh q(\tz+h) +\frac{\sigma q_x}{q(1-\sigma q_x \tz)}A(\bq)\sinh q(\tz+h) \right.\notag \\
  &+\left\{\left[1-\frac{\sigma q_x}{q(q-q_c)(1-\sigma q_x \tz)}\right]\sinh q(\tz+1)\right.\notag \\
  &-\left.\left.\left[\frac{1}{q-q_c}-\frac{\sigma q_x}{q(1-\sigma q_x \tz)}\right]\cosh q(\tz+1)\right\}\Theta(\tz+1)\right).\label{eq:v}
\end{align}
Using the continuity equation \eqref{eq:sourcepoint} then gives
\begin{align}
  u(\tz)=& \frac{\rmi q_x}{q}\left( A(\bq)\cosh q(\tz+h) -\frac{\sigma q_y^2}{qq_x(1-\sigma q_x \tz)}A(\bq)\sinh q(\tz+h) \right.\notag \\
  &+\left\{\left[1+\frac{\sigma q_y^2}{qq_x(q-q_c)(1-\sigma q_x \tz)}\right]\sinh q(\tz+1)\right.\notag \\
  &-\left.\left.\left[\frac{1}{q-q_c}+\frac{\sigma q_y^2}{qq_x(1-\sigma q_x \tz)}\right]\cosh q(\tz+1)\right\}\Theta(\tz+1)\right).
  \label{eq:u}
\end{align}

A noteworthy difference from the 2D case (\ET) is that the expressions for $u$ and $v$ have two distinct poles, each corresponding to critical layer-type solutions. Inspection of Eqs.~\eqref{eq:v} and \eqref{eq:u} reveals that, in addition to the poles for the regular wave where $\Gamma(\bq)=0$, there are critical layer type poles both at $q=q_c(\theta)$ and at $q=1/(\sigma\cos\theta\tz) = q_c/|\tz|$. 

The wave number $q_c$ was discussed above, and at length in the 2D case in \ET. 
{\color{red}
It is the wave number associated with the string of vortical flow structures generated as the source injects additional vorticity into the flow as discussed in section \ref{sec:vort}. These vortices have spatial periodicity (``wavelength'') $2\upi/q_c$ in dimensionless units.

On the other hand, $q=q_c/|\tz|$, or in dimensional terms, $\omega/k\cos \theta=U(z)$, describes a ``wave'' moving downstream instead at the same velocity as the flow at depth $\tz$, and seems likely to be caused instead by flow structures at that level. Like the critical wave, this ``wave'' (visible only in horizontal velocity components) is also only present downstream of the source.
Thus, unlike the vertical velocity $w$, the horizontal velocity components are affected in the far field by drifting flow structures from the source not only at the single depth $\tz=-1$, but at \emph{all} depths. 
}
A careful study of the resulting velocity field as well as the vorticity transport from the source into the critical layer is likely to reveal interesting and novel insights, but is a major undertaking beyond the scope of the present effort.

%%%%%%%%%%%%%%%%%%%%%%%%%%%%%%%%%%%%%%%%%%%%%%%%%%%%%%
%%%%%%%%%%%%%%%%%% S E C T I O N %%%%%%%%%%%%%%%%%%%%%
%%%%%%%%%%%%%%%%%%%%%%%%%%%%%%%%%%%%%%%%%%%%%%%%%%%%%%
\section{Summary and conclusions}

The oscillating point source in a shear flow of uniform vorticity $S$ with a free surface is analysed analytically in the framework of the Euler equation, linearized with respect to perturbation quantities. The analysis is thus a natural extension of the corresponding effort in 2D by \cite{ellingsen15a} (shorthand \ET). The submergence depth of the source is $D$, the constant water depth is $H$, and the oscillation frequency is $\omega$. The terms ``upstream'' and ``downstream'' are as seen from a system where the source and undisturbed surface are at rest.

The wave field due to the oscillating source is characterised by contributions from two distinct types of waves; the regular propagating dispersive waves which satisfy the dispersion relation in the presence of a shear current, and a set of waves resulting from the formation of a critical layer-like sub-surface flow below the surface which is advected downstream of the source. The latter ``critical wave'' typically has the form of a narrow train of waves of constant width and wavelength which remains unchanged in shape and amplitude as it is swept downstream with the background flow. In the 2D case it was straightforward to show that the critical layer corresponded to vortical flow structures akin to Kelvin's cat's eye vortices, centred at the depth of the source. A full analysis of the velocity field in the 3D case is more involved and will be considered in a future publication. 
{\color{red}
However, inspection of the vorticity equation reveals that exactly the same mechanism is at play in the three-dimensional flow as was analysed at length in \ET. The resulting thin line of undulating vorticity directly downstream of the source is quite distinct from the vorticity perturbations always present in 3D shear flow whenever waves propagate at oblique angles with the shear flow, and shift and twist the background vortex lines they pass \citep{ellingsen16}. In the present system, both vorticity perturbations are present.
}

In similar manner as for the 2D case considered in \ET, a cut-off of regular waves in some downstream directions occurs in the limit of deep water when the shear $S$ exceeds $\omega$. In 2D the result was that the downstream regular wave disappears, whereas in the 3D case all regular waves vanish for propagation directions inside a ``forbidden'' sector of propagation angles in downstream directions. When $S/\omega\to\infty$ the width of the forbidden sector approaches $\upi$, covering all directions with a downstream component. When the water depth is finite, regular waves may propagate in all directions, and for large but finite depth the ``forbidden'' sector is instead characterised by regular waves of long wavelengths and small amplitudes.

We define three nondimensional parameters to characterise the flow: the relative depth $h=H/D$, the dimensionless shear $\sigma=S/\omega$, and the (squared) Froude number $F^2 = \omega^2 D/g$. At a particular, resonant value $F^2_\text{res}(\sigma,h)$ of the Froude number, the downstream regular and critical waves resonate. A clear distinction between regular waves and critical waves can only be made at sub-resonant Froude numbers, and we provide asymptotic expressions for the far-field wave patterns in the sub-resonant case. As $F^2$ approaches the resonant value, resonance results in a wave field which grows linearly with distance from the source and which is contained within a downstream wedge symmetrical about $y=0$.

\bibliographystyle{jfm}
\bibliography{wave}

\end{document}